\documentclass[aps,pra,twoside,twocolumn,10pt]{revtex4-1}
\usepackage[colorlinks=true, citecolor=blue, urlcolor=blue ]{hyperref}
\usepackage{epsfig,newlfont,amssymb,amsfonts,amsmath,bm,subfigure,palatino,mathtools,amsthm,braket,soul,enumitem,color,graphics,graphicx,times,physics}
\usepackage[normalem]{ulem}

\begin{document}
\title{Designing refrigerators in higher dimensions using quantum spin models}

\author{Tanoy Kanti Konar$^{1}$, Srijon Ghosh$^{1}$, Amit Kumar Pal$^{2}$, Aditi Sen(De)$^{1}$}
\affiliation{$^1$Harish-Chandra Research Institute, A CI of Homi Bhabha National Institute,  Chhatnag Road, Jhunsi, Allahabad - 211019, India}
\affiliation{$^2$ Department of Physics, Indian Institute of Technology Palakkad, Palakkad 678 557, India}

\begin{abstract}
    We design quantum refrigerators based on spin-$j$ quantum $XYZ$  and bilinear-biquadratic models with individual spins attached to bosonic thermal baths. By considering both local and global master equations, we illustrate an enhancement in the performance of the refrigerators with an increase in the spin dimension irrespective of the choice of the spin models. To assess the performance of the refrigerators, we introduce a distance-based measure to quantify the local temperature of a particle with arbitrary spin quantum number $j$.   Interestingly, we find that the local temperature quantifier, defined via  minimizing the distance between a spin-$j$ thermal state and the evolved state of the spin-$j$ particle in the steady state, coincides with the population-based definition of local temperature known in the literature for spin-$\frac{1}{2}$ particles. Moreover, we demonstrate that the qualitative behavior of the distance-based local temperature  is independent of the choice of the distance measure by comparing the trace distance, Uhlmann's fidelity and relative entropy distance. We further observe by computing local  master equation that the quantum refrigerator consisting of a spin-$1/2$ and a spin-$j$ particle can lead to a lower local temperature  compared to a  refrigerator with two identical spin-$j$ particles following the $XYZ$ interactions.
\end{abstract}
\maketitle
\section{Introduction}

In the last few decades, the rapidly emerging field of quantum thermodynamics~\cite{binder2018,gemmer2004,deffner2019} has offered the prospect for unravelling fundamental laws of miniaturized quantum systems, including thermal devices such as quantum batteries~\cite{Alicki,PoliniPRL}, quantum thermal transistors~\cite{karl2016}, diodes~\cite{miranda2017} and quantum refrigerators~\cite{linden2010}. Recent progress in quantum technologies as well as in the ability of controlling quantum systems~\cite{raimond2001, bloch2008, haffner2008, pan2012} has fueled the experimental urge of constructing quantum thermal machines in order to see whether they outperform their classical counterparts. In this respect, it is also shown that quantum spin models,  implementable in different physical substrates like cold atoms~\cite{duan2003, lewenstein2007}, trapped ions~\cite{mintert2001, porras2004, monroe2021}, and nuclear magnetic resonance systems~\cite{zhang2005, rao2013, rao2014}, can serve as important platforms to realize these quantum thermal machines~\cite{modi2017,yuhan2017,modi2018,srijon2020,srijon2021, hewgill2020,konar2021}.

The main task of a quantum refrigerator consisting of a few $d$-dimensional quantum mechanical subsystems coupled with local thermal baths is to decrease the temperature of a chosen subsystem in the steady state~\cite{Allahverdyan2004,linden2010,elias2011,allahverdyan2011, clivaz2019,  arisoy2021}. This has so far been achieved using different combinations of qubits and qutrits~\cite{linden2010,naseem2020,hewgill2020, khlifi2020,bhandari2021,konar2021}. The available proposals  till date include devices in which cooling is either performed with the  help of one or more external energy source(s)~\cite{elias2011}, or in a self-contained fashion~\cite{linden2010}. To obtain the minimum possible steady-state temperature, cooling assisted by various means, such as solar energy~\cite{wang2015}, rapid measurement~\cite{erez2008}, repeated collision~\cite{dechiara2018}, periodically modulated interactions~\cite{vasco2021}, and paradigmatic quantum spin Hamiltonians~\cite{hewgill2020,konar2021} have been reported, and even achievement of Carnot efficiency in a two-qubit setup via a reverse-coupling mechanism has been shown~\cite{silva2015}. Moreover,  recent experimental  realizations of  refrigerators using trapped ions~\cite{maslennikov2019} and several experimental proposals employing superconducting qubits~\cite{hofer2016}, quantum dots~\cite{davide2013}, trapped ions~\cite{mitchison2016}, and optomechanical systems~\cite{mari2012} have made the implementation of a spin model-based fridge in laboratories a possibility.

As of now, most of the proposed and implemented quantum technologies typically involve  two-dimensional  systems due to (a) the relative ease of handling a single or a multiqubit system compared to a system involving qudits, and (b) the fact that a quantum system moves towards the classical limit with an increase in the spin quantum numbers of the constituent spins, thereby eventually losing its quantum characteristics. However, higher dimensional quantum systems are revealed to be advantageous over their lower dimensional counterparts in several quantum gadgets, including quantum key distribution~\cite{durt2003}, quantum switch~\cite{wei2019}, and quantum batteries~\cite{santos2019,dou2020,ghosh2021}, to name a few. While there has been a few attempts in constructing quantum refrigerators using constituent quantum systems with a Hilbert space dimension higher than that of a qubit or a qutrit~\cite{correa2014,wang2015,silva2016,usui2021} and even with quantum oscillator~\cite{levy2012}, to the best of our knowledge, realization of quantum refrigerators using quantum spin models constituted of particles with arbitrary spin-quantum number remains an unexplored area, which we address in this work. 

Our design of the quantum refrigerator bears two distinct features. (a) We employ interacting quantum spin systems with nearest-neighbor interactions, namely the quantum $XYZ$ model~\cite{qptbook1} and the  bilinear-biquadratic  (BB) model~\cite{Sutherland1975, Takhtajan1982, Babu82, Fath91, Fath93}, consisting of two or three spins having spin quantum number $j$.  (b) In order to quantify the performance of the refrigerator, we introduce a definition of local temperature for a spin-$j$ system that uses the minimum distance between the time-evolved state of the system and a canonical thermal state. We prove that the introduced measure reduces to the measure of local temperature based on population of the ground state in case of spin-$1/2$ systems that is already available in literature~\cite{linden2010}. We further demonstrate that the proposed definition of local temperature is independent of the choice of the distance measure, by considering the trace distance, the relative entropy distance \cite{nielsenchuang} and Uhlmann's fidelity ~\cite{uhlmann1976}.

We derive the explicit forms of the Lindblad operators corresponding to subsystems with spin quantum number $j$, when the quantum master equation is constructed following a local approach. We show, by solving the local quantum master equation, that a two- and a three-spin system of identical spins with spin-$j$ governed by the $XYZ$ type, or the bilinear-biquadratic interactions and connected to local bosonic thermal baths can serve as a refrigerator for a chosen spin in the system. The performance of the refrigerator, as quantified by the proposed distance-based local temperature and the normalized von Neumann entropy of the steady state of the cold spin, exhibits diminution with an increase in $j$, demonstrating the dimensional advantage. We also show that a duo of a spin-$\frac{1}{2}$ and a spin-$j$ system cooling the spin-$\frac{1}{2}$ particle has better dimensional benefit compared to a system of two identical spin-$j$ particles with local master equation. The dimensional improvement is found to persist even when one considers the global approach of constructing the quantum master equation.

The paper is organised as follows.  The setup of the refrigerator with spin models, local thermal baths, and their interactions, as well as the derivation of Lindblad operators for local quantum master equations  are described in Sec. \ref{sec:model}. In Sec. \ref{sec:definition_temperature},  we introduce the concept of quantifying local temperature using distance measures, and prove that it coincides with the population-based definition of local temperature for spin-$1/2$ systems. We also discuss the use of von Neumann entropy as an indicator for local temperature. The performance of the refrigerators constituted of two spins using these figures of merit  is reported in Sec. \ref{sec:twospinR}. The analysis on the  refrigerator with three spin-$j$ particles is carried out in Sec. \ref{sec:threespinR}, while we conclude in Sec. \ref{sec:conclu}.

\section{Design for Quantum Refrigerator}
\label{sec:model}

In this section, we briefly discuss the system-environment setup, a part of which acts as a quantum refrigerator for the rest under specific conditions of the system as well as the system-environment interaction parameters. For  local master equation, we also derive the Lindblad operators applicable in higher dimensions.

\begin{figure}
    \centering
    \includegraphics[width=\linewidth]{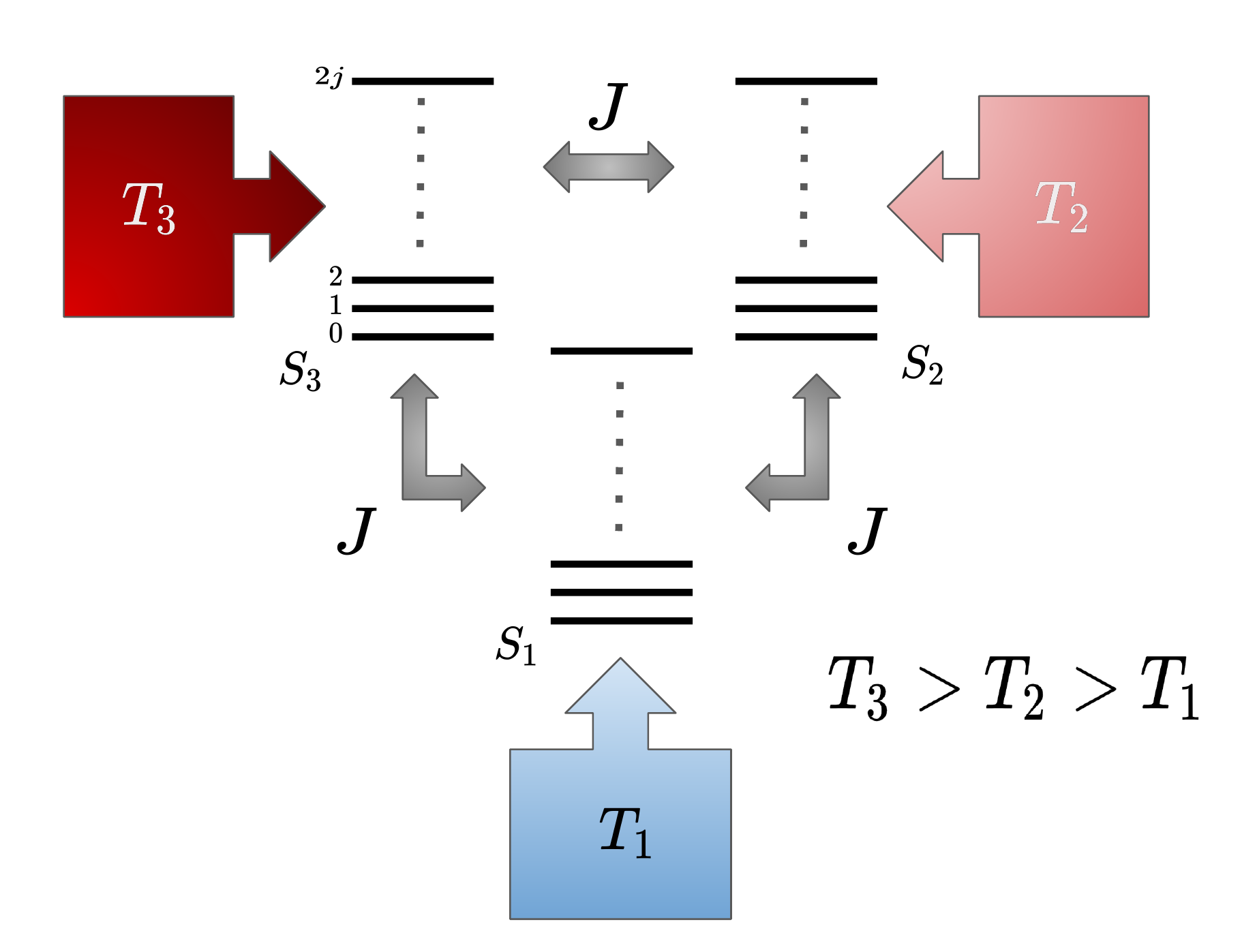}
    \caption{(Color Online.) \textbf{Schematic representation of a quantum refrigerator consisting of three spin-$j$ particles.} Each of the spin is coupled with  non-interacting thermal baths having temperature $T_{1}, T_{2}$ and $T_{3}$. $J$ is the interaction strength between the spins.}
    \label{fig:schematic}
\end{figure}

\subsection{Small spin clusters as system}

We use small clusters of particles with specific types of two-body interactions to design the refrigerator, where the individual particles can take half-integer as well as integer spins. The total Hamiltonian $H_{sys} = H_{loc} + H_{int}$ of the system consists of two parts -- (a) the local Hamiltonian $H_{loc}$ given by 
\begin{eqnarray}
H_{loc}=\sum_{r=1}^{N}h_{r}S^{z}_{r},
\label{eq:h_local}
\end{eqnarray}
and (b) the interaction Hamiltonian $H_{int}$ governed by the spin-spin interactions. We particularly focus on two types of spin-spin interactions, namely, the nearest neighbor $XYZ$  interaction giving rise to the  Hamiltonian~\cite{qptbook1} 
\begin{eqnarray}
H_{xyz}&=&J\sum_{r=1}^{N}\left[(1+\gamma)S^{x}_{r}S^{x}_{r+1}+(1-\gamma)S^{y}_{r}S^{y}_{r+1}\right]\nonumber \\
&&+J\Delta\sum_{r=1}^NS_r^zS_{r+1}^z,
\label{eq:hint_xy}
\end{eqnarray}
and the bilinear-biquadratic Hamiltonian~\cite{Sutherland1975, Takhtajan1982, Babu82, Fath91, Fath93}
\begin{eqnarray}
\label{eq:hint_bb} 
H_{B}(\phi)&=&J\cos \phi\sum_{r=1}^{N}\vec{S}_r\vdot\vec{S}_{r+1}+J\sin \phi\sum_{r=1}^{N}(\vec{S}_r\vdot\vec{S}_{r+1})^2,\nonumber \\ 
\end{eqnarray}
where we assume periodic boundary conditions unless otherwise mentioned. 
Here, $S^{\nu}_{r}$ ($\nu = x, y, z$) are the $(2j + 1)$ dimensional spin matrices (for a spin-$j$ particle, \(j=\frac{1}{2},1,\frac{3}{2},...\) for the $XYZ$ model, and \(j=1,\frac{3}{2}...\) for the $BB$ Hamiltonian) acting on the site $r$, and $N$ is the total number of particles in the system. The $r$th spin in the system is subject to a magnetic field of strength $h_{r}$ in the $z$-direction. When the interaction is of the $XYZ$ type, $J$ is the strength of the spin-spin interaction strength between the nearest-neighbor spins,  while  $\gamma$ and $\Delta$ represent the $xy$- and the $z$-anisotropy parameters respectively. When \(\Delta = 0\) and \(\gamma=0\), the Hamiltonian represents the \(XX\) model while \(\gamma =0\), and \( \Delta \ne 0 \) gives the $XXZ$ mmodel. On the other hand,  $J\cos \phi$ and $J\sin \phi$  are the interaction strengths for  the linear and the quadratic terms in the BB  Hamiltonian respectively, where the parameter $\phi$ governs the phases of the system in the absence of the local magnetic field~\cite{Sutherland1975, Takhtajan1982, Babu82, Fath91, Fath93}. With the aim of designing small quantum thermal machines and investigating the effect of a change in 
the Hilbert space dimension of the system on the performance of the machine, we typically restrict the values of $N$ to be $N=2$ and $N=3$. 

Note that in Eqs.~(\ref{eq:hint_xy}) and (\ref{eq:hint_bb}), we  assume all the particles in the interacting quantum spin model to have identical spin value $j$. However, in a more general situation, one may consider different spin values for different particles at different lattice sites. An example of such cases will be discussed in Sec.~\ref{subsec:mixed_spin} for $N=2$.

\subsection{Interaction with the local Bosonic baths}
\label{subsec:osd}

Consider a situation where each spin $r$ in the system is interacting with a local bath $B_{r}$, $r = 1,2, ..., N$ (see Fig.~\ref{fig:schematic} for an illustration with $N=3$), which is a collection of harmonic oscillators, described by the bath Hamiltonian 
\begin{equation}
    H_{B_r} = \sum_{\eta=0}^{\eta_{c}}  h_r a_{\eta}^{\dagger} a_{\eta},
    \label{bath_eqn.}
\end{equation}
so that the total bath Hamiltonian for $N$ baths is given by $H_B=\sum_{r=1}^N H_{B_r}$. Here,  $a_{\eta}$ ($a_{\eta}^{\dagger}$) is the annihilation (creation) operator of the mode $\eta$, such that $[a_{\eta} , a_{\tilde{\eta}}^{\dagger}] = \delta(\eta - \tilde{\eta})$, and $\eta_{c}$ is the cut-off frequency of the bath. We consider the absolute temperature of the bath $B_r$ to be $T_r^0$, and the baths are local in the sense that the bath $B_r$ affects only the spin $r$ in the entire bath-environment setup. The Hamiltonian defining the interaction between the systems and baths, denoted by $H_{SB}$, reads as
\begin{equation}
   H_{SB}=\sum_{r=1}^N\sum_{\eta}\left(S_r^+\otimes a_\eta+S_r^-\otimes a_\eta^\dagger\right) 
   \label{H_sb},
\end{equation}
where $S_r^+$ ($S_r^-$) is the spin raising (lowering) operators of $r$th spin, given by $S_r^\pm=S^x_r\pm\text{i}S^y_r$.

We consider a scenario where at $t=0$, $H_{sys}=H_{loc}$, such that each spin is in thermal equilibrium with its respective bath and has an initial temperature equal to the bath temperature $T_r^0$, such that the initial ($t=0$) state of the $r$th spin is represented by a diagonal density matrix $\rho_r^0$. In the eigenbasis of $S_r^{z}$ having eigenvalues $j,j-1\ldots -j$, it takes the form as
\begin{eqnarray}
\rho_r^0&=&\tau_{r}^{2j}(0)\dyad{2j}{2j}+\tau_{r}^{2j-1}(0)\dyad{2j-1}{2j-1}\nonumber\\ &&+\cdots+\tau_{r}^0(0)\dyad{0}{0},  
\end{eqnarray}
where 
\begin{eqnarray} 
\tau_{r}^\mu(0)=\frac{\exp(-\mu\beta_r^0 h_r)}{\sum_{\mu=0}^{2j}\exp(-\mu\beta_r^0 h_r)}
\end{eqnarray}
with $\sum_{\mu=0}^{2j}\tau_{r}^\mu(0)=1$, such that the initial state of the $N$-spin system is $\rho(0) = \bigotimes_{r = 1}^{N} \rho_{r}^0$, and $\beta_r^0=1/k_BT_r^0$, \(k_B\) being the Boltzmann constant, which is set to \(1\). After the interaction Hamiltonian $H_{int}$ is turned on at $t>0$ such that $H_{sys}=H_{loc}+H_{int}$ at $t>0$, the time-dynamics of the system is governed by the quantum master equation (QME)~\cite{Petruccione}, given by  
\begin{equation}
\dot\rho=-\text{i}[H_{sys},\rho]+\mathcal{L}(\rho),
\label{eq:qme}
\end{equation}
where $\mathcal{L}(.)$ represents the dissipator, emerging out of the spin-bath interactions. The solution of Eq.~(\ref{eq:qme}) provides the state, $\rho(t)$, of the system as a function of $t$.

\subsection{Dissipators: local vs. global}

There exists two competing approaches to determine the dissipator in the QME (Eq.~(\ref{eq:qme})) -- (a) a global approach, where transitions between the eigenstates of the entire system represented by the Hamiltonian $H_{sys}$ is considered, and (b) a local treatment of the quantum master equation, considering only the transitions between the eigenstates of the individual subsystems labeled by $r=1,2,\cdots,N$.  In the former, the dissipator $\mathcal{L}(\rho)=\sum_{r=1}^NL_r(\rho)$, with 
\begin{widetext}
\begin{eqnarray}
L_r(\rho) &=& \sum_{\omega}\gamma_r(\omega) \left[A_r(\omega)\rho A_r^\dagger(\omega)-\frac{1}{2}\left\{A_r^\dagger(\omega)A_r(\omega),\rho\right\}\right],
\label{lindblad_global}
\end{eqnarray}
\end{widetext}
where the operators $A_r(\omega)$ are the Lindblad operators corresponding to the $r$th spin for a transition amounting energy $\omega$ among the energy levels of the system, defined by the equation  
\begin{eqnarray}
\text{e}^{\text{i}H_{sys}t}\left(S_r^++S_r^-\right)\text{e}^{-\text{i}H_{sys}t}=2\sum_{\omega}A_r(\omega)\text{e}^{-\text{i}\omega t}. 
\end{eqnarray}
Explicit forms of $A_r(\omega)$ can be derived by decomposing the spin-part of the system-bath interaction Hamiltonian in the eigenbasis of $H_{sys}$, and may not always be analytically derivable in cases of complex Hamiltonians with large number of spins. The transition rate, represented by $\gamma_r(\omega)$, corresponds to the jump through an energy gap $\omega$ for the spin $r$, and depends on the spectral function and the cut-off frequency of the bath. For baths with an Ohmic spectral function $\kappa_r(\omega)=\left[\exp(\beta_r\omega)-1\right]^{-1}$ and a cut-off frequency $\omega_c$ which are the same across baths,  
\begin{eqnarray}
\gamma_r(\omega) &=& f_r(\omega)[1+\kappa_r(\omega)],\text{ for }\omega\geq 0,\nonumber \\
\gamma_r(\omega) &=& f_r(|\omega|)\kappa_r(|\omega|), \text{ for }\omega<0,
\end{eqnarray}
with $f_r(\omega)=\alpha_r\omega\exp(-\omega/\omega_c)$, and $\alpha_r$ being a constant for the $r$th bath, representing the spin-bath interaction strength. Under Markovian approximation, $\max\{\alpha_r\}\ll 1$.

Note here that applicability of the local and global approaches to derive master equations  is not fully settled in literature~\cite{Levy_2014}. There are proposals for both methods being self-contained having their own regimes of validity~\cite{Hofer_2017,dechiara2018}. In this paper, we use local approach where \(h_i\gg J_{i,i+1}\), and global approach is considered in the region \(h_i\simeq J_{i,i+1}\).  While the global approach of analytically determining the master equation can be challenging, the explicit form of the Lindblad operators can, however, be determined if one takes the local one. Note that in the limit where the spin-spin and the spin-bath interaction strengths are so small that  $H_{sys}\approx H_{loc}$ $\left(\gamma_r(\omega),J<<h_r\right)$, one may calculate the  Lindblad operators using the eigenstates of $H_{loc}$ only, which we present in the following for the systems consisting of  spin-$j$ particles.   

\noindent\textbf{Lindblad operators for spin-$j$ local QME.} Let us demonstrate the explicit form of the Lindblad operators considering a system of two spin-$j$ particles ($r=1,2$), subject to magnetic fields of strength $h_1$ and $h_2$ in the $z$-direction. Eigenvalues of the local Hamiltonian (Eq.~(\ref{eq:h_local})) for this system can be written as $(-j+a)h_1+(-j+b)h_2$,  corresponding to eigenstates $\ket{a}\otimes \ket{b}$, where $a,b\in[0,2j]$. Writing the system-bath interaction Hamiltonian as    
\begin{equation}
    H_{SB}=\sum_{r=1}^2\sum_{\eta}\left[S_r^x\otimes(a_\eta+a_\eta^\dagger)+S^y_r\otimes \text{i}(a_\eta-a_\eta^\dagger)\right],
\end{equation}
the Lindblad operators corresponding to the $r$th bath can be determined as~\cite{Petruccione} 
\begin{equation}
    A_r(\omega)=\sum_{\epsilon_q-\epsilon_p=\omega}\dyad{\epsilon_p}{\epsilon_p}S_r^x\dyad{\epsilon_q}{\epsilon_q},
\end{equation}
where $\epsilon_p$ ($\epsilon_q$) is the $p$th ($q$th) eigenstates of $H_{loc}$. Performing algebra for spin-$j$ particles corresponding to the first spin leads to
\begin{eqnarray}
A_1(\omega)&=&\sum_{\epsilon_q-\epsilon_p=\omega}\frac{1}{2}[\sqrt{j(j+1)-a(a+1)}\ket{a^\prime}\bra{a}\delta_{a^\prime,a+1}\nonumber\\
&&+\sqrt{j(j+1)-a(a-1)}\ket{a^\prime}\bra{a}\delta_{a^\prime,a-1}]\otimes\ket{b}\bra{b},\nonumber \\
\end{eqnarray}
where $A_2(\omega)$ can also be determined with similar calculation. Considering (a) only transitions with positive $\omega$ implying $\omega=h_1$, and noticing that (b) non-zero matrix elements for $A_1(\omega)$ requires transitions to be within consecutive energy levels only, $a\in[1,2j]$, $b\in[0,2j]$, and $a^\prime=a-1$. Therefore,   the desired Lindblad operator can be represented as
\begin{eqnarray}
\nonumber A_1(\omega)&=&\frac{1}{2}\left[\sum_{a=1}^{2j}\sqrt{j(j+1)-a(a-1)}\dyad{a-1}{a}\right]\nonumber\\
&&\otimes\left[\sum_{b=0}^{2j}\dyad{b}{b}\right]\nonumber\\
&=&\frac{1}{2}(S^{-}_1\otimes\mathbb{I}),
\end{eqnarray}
where $\mathbb{I}$ is the identity operator in the Hilbert space of a spin-$j$ particle. This calculation can be extended to a system of $N$ spin-$j$ particles also, where the Lindblad operators for the $r$th bath are given by 
\begin{eqnarray}
A_r(\omega)&=&\frac{1}{2} \mathbb{I} \otimes \cdots S_r^- \otimes \ldots \mathbb{I},\nonumber\\  A_r^\dagger(\omega)&=&\frac{1}{2}\mathbb{I} \otimes \cdots S_r^+ \otimes \ldots \mathbb{I}. 
\label{eq:local_lindblad}
\end{eqnarray}
From here on, we discard the $\otimes$ (tensor product) in the expressions for Lindblad operators for brevity.



Note that in the case of the local approach, the time-evolution of the state of a system of identical spins depends on the initial state of the system, as presented in the following proposition. \\

\noindent\textbf{Proposition I}. \emph{For a system of identical spins governed by a \(XX\) Hamiltonian, if 
\begin{eqnarray}
\frac{h_r}{T_r^0}=\text{constant} \quad \forall\; r=1,2,\cdots,N,
\label{eq:no_evolution}
\end{eqnarray}
the system does not evolve with time as long as the dynamics is governed by a local quantum master equation.}
\begin{proof}
To prove it, we first note that $\rho(t+\delta t)$, with \(\delta t\) being the small increment in time, can be expanded as  
\begin{eqnarray}
\rho(t+\delta t) &=& \rho(t)+\eval{\pdv{\rho}{t}}_t\delta t+ \mathcal{O}(\delta t^2).
\end{eqnarray}
Performing the expansion about $t=0$, and neglecting higher order terms, we obtain
\begin{eqnarray}
\rho(\delta t)&\approx&\rho(0)+\eval{\pdv{\rho}{t}}_{t=0}\delta t \nonumber \\ &&=\rho(0)-\text{i}[H_{sys},\rho(0)]-\mathcal{L}(\rho(0)). 
\end{eqnarray}
Since the interaction between the spins is absent at $t=0$, for dissipators constructed of Lindblad operators of the form given in Eq.~(\ref{eq:local_lindblad}), $\mathcal{L}(\rho(0))=0$. Also, at $t=0$, the condition in Eq.~(\ref{eq:no_evolution}) suggests identical initial states for all spins in the system, implying $[H_{sys},\rho(0)]=0$ (see Appendix~\ref{prop_xx}), leading to $\eval{\pdv{\rho}{t}}_{t=0}=0$. This can be continued for an arbitrary small time increment $\delta t$ such that $t=n\delta t$, $n$ being an integer, when $\rho(t)=\rho(n\delta t)=\rho(0)$. Therefore, there is no time-evolution of the state in the system and hence the proof.
\end{proof}

\section{Quantifying local temperature in higher dimension}
\label{sec:definition_temperature}

To assess the performance of the quantum refrigerator described in Sec, \ref{sec:model}, we consider the amount of \emph{local cooling} in one of the chosen spins, and to quantify this, we propose a definition of local temperature which remains valid for arbitrary spins. For such loal cooling, different notions exist in literature, eg., increase of ground state population of the cold subsystem~\cite{linden2010,deAssis2019}, or energy flow from cold bath to the subsystem~\cite{levy2012}. Note that defining local temperature using the ground-state population, in the same vein as done in, for example,~\cite{linden2010}, for qubit systems is not possible here due to the non-uniqueness of the temperature defined in this fashion. Motivated by the logical question as to whether the increase in the ground state population can at all be exploited in any other way to quantify the local temperature of a qudit,  in the following, we define the distance-based measure for estimating temperature of the qudit.

In this paper, we shall focus on the scenario where one aims to cool a chosen spin in the spin-bath setup during the dynamics. The $r$th spin in the system is said to achieve a \emph{local steady-state cooling} by virtue of the dynamics of the system if and only if $T_r^0>T_r^s$, where $T_r^s=T_r(t\rightarrow\infty)$ is the local steady-state temperature of the $r$th spin. Note that the chosen spin-bath interaction (see Sec.~\ref{subsec:osd}) ensures a diagonal reduced density matrix, 
\begin{eqnarray}
\rho_r(t)=\text{Tr}_{\underset{j,k=1,2,3}{j,k (\neq r)}}\left[\rho(t))\right], \, r=1,2,3, 
\end{eqnarray} 
for the $r$th spin. In the case of a system having a spin-$\frac{1}{2}$ particle at the $r$th site, we know that $\rho_r(t)$ takes the form   as
\begin{eqnarray}
\rho_r(t)&=&\tau_{r}^0(t)\dyad{0}{0}+\tau_{r}^1(t)\dyad{1}{1}, 
\label{eq:local_state_spin-half}
\end{eqnarray}
where $\tau_r^0(t)$ $(\tau_r^1(t))$ can be identified as the time-dependent population of the state $\ket{0}$ ($\ket{1}$), and can be used to define a \emph{population-based local temperature} (PLT) for the $r$th spin as a function of time, given by 
\begin{eqnarray}
T_r(t)=\frac{h_r}{\ln\left[{\tau_r^1(t)}^{-1}-1\right]}. 
\end{eqnarray}
While this protocol for defining the local temperature of a spin-$\frac{1}{2}$ particle is well-established~\cite{linden2010}, the definition of a local temperature remains a scarcely explored area with little literature~\cite{burke2021} (cf. \cite{silva2016, usui2021}),  when the subsystems have Hilbert space dimension larger than $2$ (eg. spin-$j$ particles). In the latter situation, the local density matrix of the $r$th spin, having  the form 
\begin{eqnarray}
\rho_r(t)&=&\tau_{r}^{2j}(t)\dyad{2j}{2j}+\tau_{r}^{{2j-1}}(t)\dyad{2j-1}{2j-1}\nonumber\\ &&+\cdots+\tau_{r}^{{0}}(t)\dyad{0}{0}, 
\label{eq:local_state_time_dependent}
\end{eqnarray}
depends on a total of $2j$ parameters, and hence defining a \emph{unique} local temperature for the $r$th spin-$j$ particle following the protocol for spin-$\frac{1}{2}$ particles~\cite{linden2010} is not possible. In this regard, we put forward a distance-based quantifier for local temperature in the subsequent subsections, and justify its importance in the context of investigating performance of a quantum refrigerator constructed out of quantum spin models. In this way, a set of definition for local temperature emerges depending on the choice of valid distance measures although we show that they qualitatively behave  in a similar fashion.

\subsection{Estimating local temperature using distance measures}
\label{subsec:local_cooling}

Let us consider an arbitrary canonical thermal state of the $r$th spin-$j$ particle in the system, having an absolute temperature $T_r^\prime$. The canonical thermal state is given by 
\begin{equation}
    \tilde{\rho}_r=\frac{\exp(-\beta^\prime_r h_r S^z_r)}{\text{Tr}\left[\exp(-\beta^{\prime}_r h_r S^z_r)\right]},
    \label{trial}
\end{equation}
where $\beta^\prime_r=1/k_BT^\prime_r$ is the inverse temperature, and $h_{r}$ is the strength of the external magnetic field of the $r$th spin. We define the \emph{distance-based local temperature} (DLT), $T_r^D(t)$, for the $r$th spin described by the steady state $\rho_r(t)$ obtained via the dynamics (see Eq.~ (\ref{eq:local_state_time_dependent})) as
\begin{eqnarray}
T_r^D(t)=\frac{1}{\underset{T_r^{\prime}}{\min} D(\tilde{\rho}_r,\rho_r(t))}, 
\label{eq:loc_temp}
\end{eqnarray}
where $D(\tilde{\sigma},\sigma)$ is an appropriate distance measure between the density matrices $\tilde{\sigma}$ and $\sigma$.  There exists a number of distance measures in literature, including the trace distance~\cite{nielsenchuang}, the Hilbert-Schmidt distance~\cite{Ozawa2000}, Uhlmann fidelity~\cite{uhlmann1976, jozsa1994}, and the relative entropy distance~\cite{mendon2008} to name a few, which can be used to quantify the local temperature, and the use of a particular measure may depend on specific situations. In the following sections, we shall compare the performances of different distance measures in context of faithfully quantifying the local temperature of a spin-$j$ system. In order to justify the importance of such a definition of local temperature, we present the following proposition.

\noindent\textbf{$\blacksquare$ Proposition II.} \emph{For a spin-$\frac{1}{2}$ particle, the distance-based local temperature is equivalent to the population-based local temperature at all times, when trace distance is chosen as the distance measure.}

\begin{proof}
Let us define  
\begin{eqnarray}
y_r=\frac{\exp(-\beta^\prime_rh_r/2)}{\exp(-\beta^\prime_rh_r/2)+\exp(\beta^\prime_rh_r/2)},
\end{eqnarray}
and write $D(\tilde{\rho}_r,\rho_r(t))$, at an arbitrarily fixed time instant $t$, as a function of $y_r$ as
\begin{eqnarray}
 D(y_r)&=& \frac{1}{2}\text{Tr}\sqrt{(\tilde{\rho}_r-\rho_r(t))^\dagger(\tilde{\rho}_r-\rho_r(t))} \nonumber \\
 &=&\frac{1}{2}\left(\left|y_r-\tau_r^1(t)\right|+\left|(1-y_r)-\tau_r^0(t)\right|\right),
\end{eqnarray}
which, using $\tau_{r}^0=1-\tau_r^1$, becomes  
\begin{eqnarray}
 D(y_r)&=&\frac{1}{2}\left(\left|y_r-\tau_r^1(t)\right|+\left|\tau_r^1-y_r\right|\right). 
\end{eqnarray}
Since $D(y_r)\geq 0$ by virtue of being a distance measure, and $D(y_r)=0$ for $y_r=\tau_r^1$, the DLT is obtained from the equation $y_r=\tau_r^1$ by solving for $T_r^\prime$ as 
\begin{eqnarray}
T_r^\prime=\frac{h_r}{\ln\left[{\tau_r^1(t)}^{-1}-1\right]}. 
\label{eq:dlt-plt}
\end{eqnarray}
Since Eq.~(\ref{eq:dlt-plt}) holds for an arbitrary $t$,  the DLT is equivalent to the PLT at all times. Hence the proof. \end{proof}

In the subsequent sections, we shall discuss the steady-state cooling of a spin in the system using DLT as a quantifier for cooling, where for brevity, we denote the steady-state DLT as $T_r^{s}=T_{r}^D(t\rightarrow\infty)$.

\begin{figure*}
    \centering
    \includegraphics[width=\textwidth]{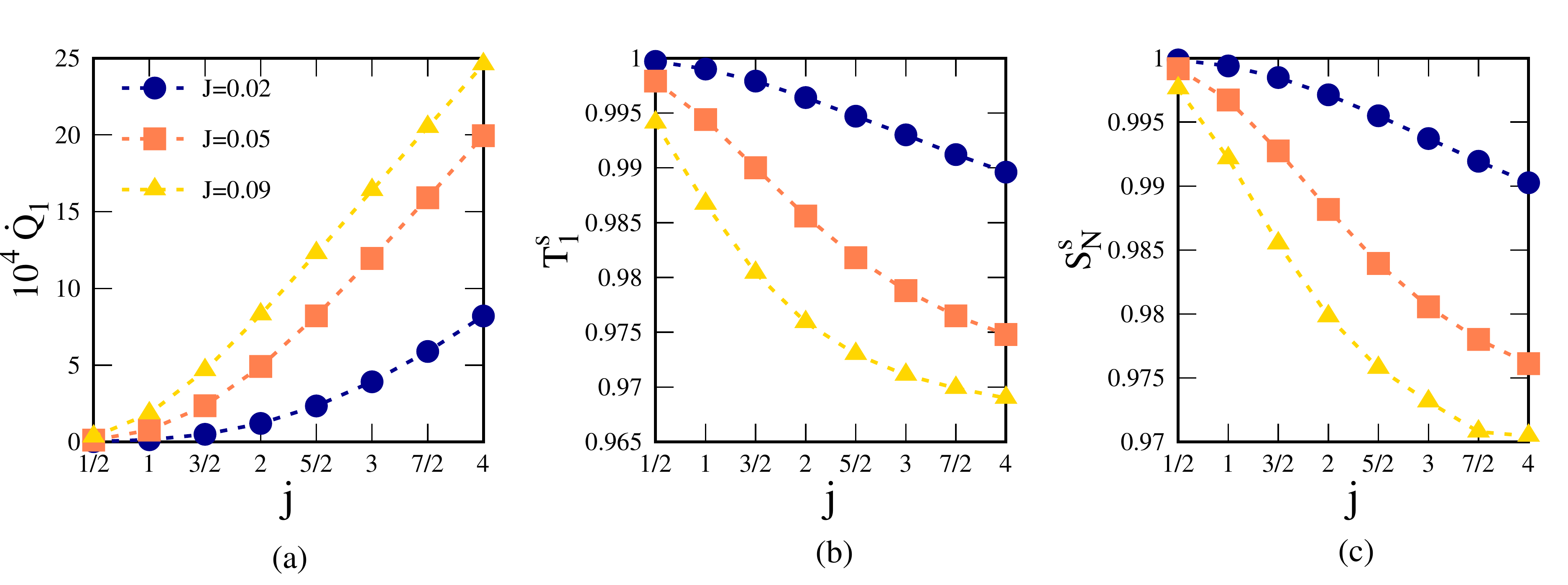}
    \caption{(Color online.) \textbf{Variation of (a) heat current, (b) steady-state temperature and (c) normalized von-Neumann entropy (vertical axis) for a two-spin system as functions of the  dimension of the first subsystem, \(j\) (horizontal axis).} The refrigerator consists of two  identical spins which are interacting according to the $XX$ Hamiltonian given in Eq. (\ref{eq:hint_xy}) with \(\gamma =0\) and \(J \Delta =0\). We compute these quantities by solving local QME. Circles, squares and triangles represent different interaction strengths, namely $J = 0.02$, $J = 0.05$ and $J = 0.09$ respectively.  The local external magnetic fields  of the first and the second spins are respectively $h_1=1.1$, and $h_2=1.3$ while the corresponding initial temperature of the first and the second  spins are $T_1(0)=1$ and $T_2(0)=1.1$ respectively.  Here the spin-bath interaction is chosen as $\Gamma=0.05$.  The dimensional advantage according to the figures of merit for the refrigerator is clearly visible.  All the axes are dimensionless. }
    \label{fig:xyz_two_spin_local}
\end{figure*}

\begin{figure*}
    \centering
    \includegraphics[scale=0.45]{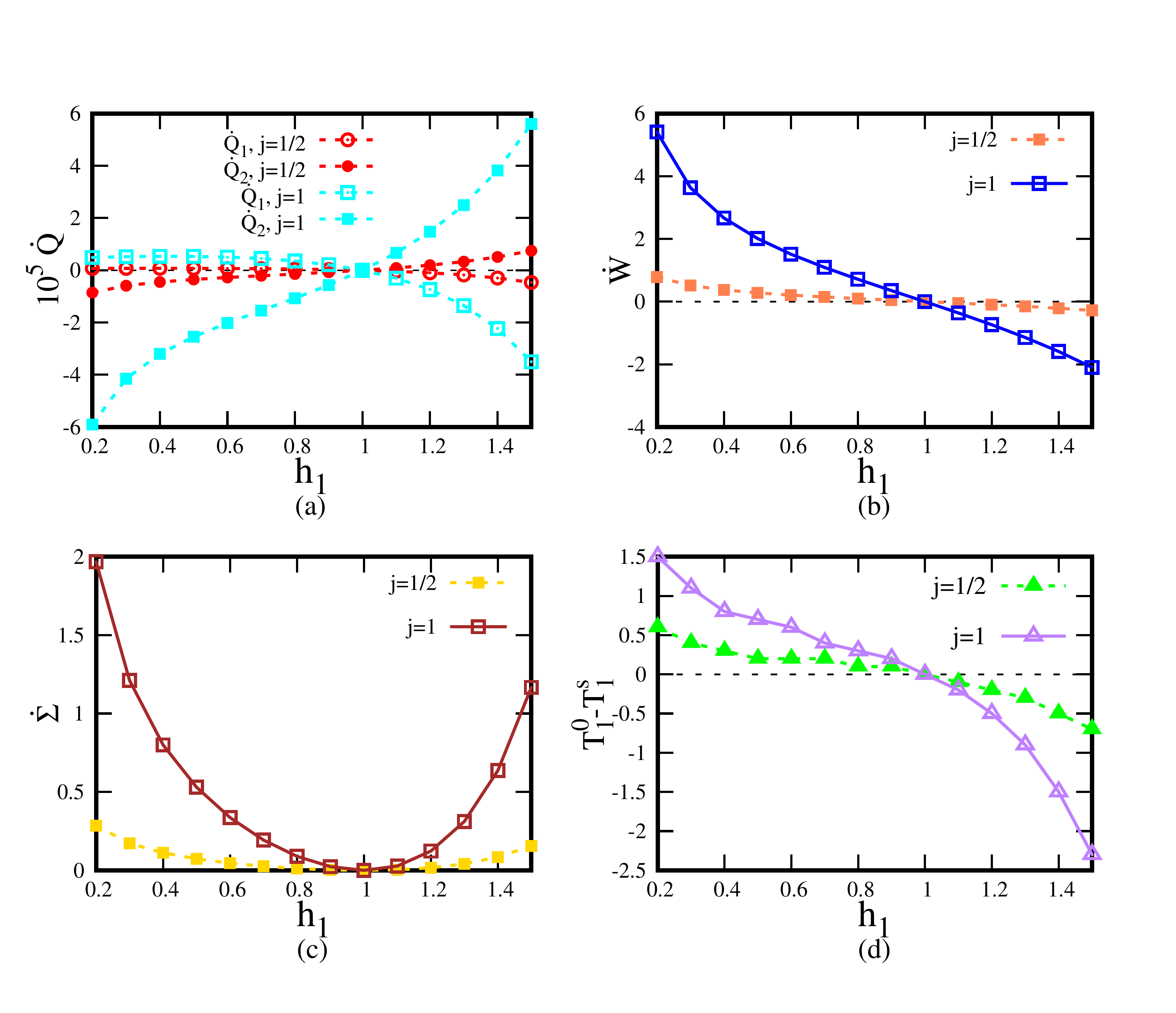}
    \caption{(Color online.) \textbf{Variation of (a) heat current, (b) work, (c) entropy production rate, and (d) steady-state temperature (vertical axis) with  initial magnetic field of the first subsystem, \(h_1\) (horizontal axis) in the case of two-qudit systems, where both are either spin-$1/2$, or spin-$1$ particles.} The refrigerator consists of two  identical spins which are interacting according to the $XX$ Hamiltonian with \(\gamma =0\) and \(J \Delta =0\) for spin-$1/2$ and spin-$1$ systemS.  The local external magnetic fields  of the second spin is $h_2=2.4$, while the corresponding initial temperature of the first and the second  spins are $T_1(0)=1$ and $T_2(0)=2.4$ respectively.  Here the spin-bath interaction strength is chosen as $\Gamma=0.05$.  All the axes in all figures are dimensionless.}
    \label{fig:xyz_two_spin_validity}
\end{figure*}

\subsection{Entropy-based estimation of local temperature}

In situations where a spin-$j$ subsystem of a quantum spin model in a system-bath setup described in Sec.~\ref{sec:model} attains a local steady-state cooling, the entropy of the subsystem in the steady state should  be lower than the initial entropy of the subsystem at $t=0$, providing a signature of the cooling phenomena. In order to carry out a quantitative investigation, we define an entropy-based estimated temperature, quantified by the normalized entropy for the steady state, as  
\begin{equation}
    S_N^s = \frac{S(\rho_{r}(t \rightarrow \infty))}{S(\rho_{r}(0))},
    \label{normalized_entropy}
\end{equation}
where $S(\rho)=-\text{Tr}(\rho\log_2\rho)$ is the von Neumann entropy for the density matrix $\rho$. A local steady-state cooling of the $r$th spin is indicated by $S_N^s <1$ while its positivity implies heating. The qualitative variations of $S_N^s$ as functions of the relevant system parameters as well as with  increasing dimension of the Hilbert spaces of the subsystems are similar to those for DLT, as we shall demonstrate in the subsequent sections. 

\subsection{Local heat current}
\label{subsec:local_heat_current}

An important quantity, providing the indication as to whether an $N$-spin system is operating as a refrigerator for the $r$th spin, is the local heat current at the steady state, defined as~\cite{Petruccione} 
\begin{eqnarray}
\dot{Q}_r=\text{Tr}\left[H_{sys}\mathcal{L}_r(\rho^s)\right],
\label{eq:global_heat_current}
\end{eqnarray}
where $\rho^s$ is the steady state $\rho(t\rightarrow\infty)$ of the entire system. A positive value of $\dot{Q}_r$ represents a situation where heat flows from the bath $B_r$ to the $r$th spin in the steady state, which is at a lower temperature $T_r^s<T_r^0$ if a steady-state cooling has been achieved. The value of $\dot{Q}_r$, therefore, is expected to be positive in accordance with a cooling indicated by $T_r^{s}$ and $S_N^s$.

At this point, a comment on the applicability of the local and global QMEs is in order. The proper way to characterize refrigeration is through the heat current, and appropriately defining the form of ``heat" considering the evolution of the system by local and global master equations is a delicate issue~\cite{Wichterich2007,landi_rmp_2022}. For a consistent definition, in the chosen parameter space, the solution of  the applied equation has to be consistent with the laws of thermodynamics which dictates the conservation of energy and the direction of the spontaneous flow of heat. The validity of a certain thermodynamic process depends on the entropy production rate, 
\begin{eqnarray}
\dot{\Sigma}=\dot{S}-\sum_{r}\frac{\dot{Q}_r}{T_r}\ge 0,
\label{eq:entropy_production_rate}
\end{eqnarray}
which validates the second law of thermodynamics \cite{landi_rmp_2022}. Here,  $T_r$ denotes the temperature of the $r$-th bath, $\dot{S}$ denotes the change of entropy denoted as \(-\sum_{r}\Tr{L_r(\rho)\log{\rho}}\), and $\dot{Q}_r$ is the heat current flowing to and fro between the system and the environment (see Eq.~(\ref{eq:global_heat_current})) and the corresponding work rate denoted by $\dot{W}=\mbox{Tr}[\{\sum_r\mathcal{L}_r(\rho_s)\}H_{int}]$. Note that the definition of $H_{sys}$ in Eq.~(\ref{eq:global_heat_current}) may vary depending on the choice of a local, or a global approach to define the QME, and an inappropriate choice of the QME may lead to anomalous values of $\dot{Q}_r$, although a steady-state cooling for the $r$th spin may be  indicated by the values of $T_r^{s}$ and $S_N^s$~\cite{ghoshal2021,konar2021}. While $H_{sys}$ is the total system Hamiltonian in the case of a global QME, corresponding to the  local QME~\cite{landi_rmp_2022},  
\begin{eqnarray}
\dot{Q}_r=\Tr\left [H_{loc}^r\mathcal{L}_r(\rho_s)\right ].
\end{eqnarray}
We further point out here that a valid thermodynamic process must obey the energy balance equations. There are two types of such balance equations used in literature~\cite{barra2015,dechiara2018}, dealing with (i) energy conservation, and (ii) the entropy production. For the setup considered in this paper,  validity of detailed balance in local QME is ensured by \([H_{loc}+H_B,H_{SB}]=0\)~\cite{dechiara2018}, while we elaborate on the validity of the entropy production rate equation in the subsequent sections as we discuss specific constructions of small refrigerators in a case-by-case basis.

 \begin{figure*}
    \centering
    \includegraphics[width=\textwidth]{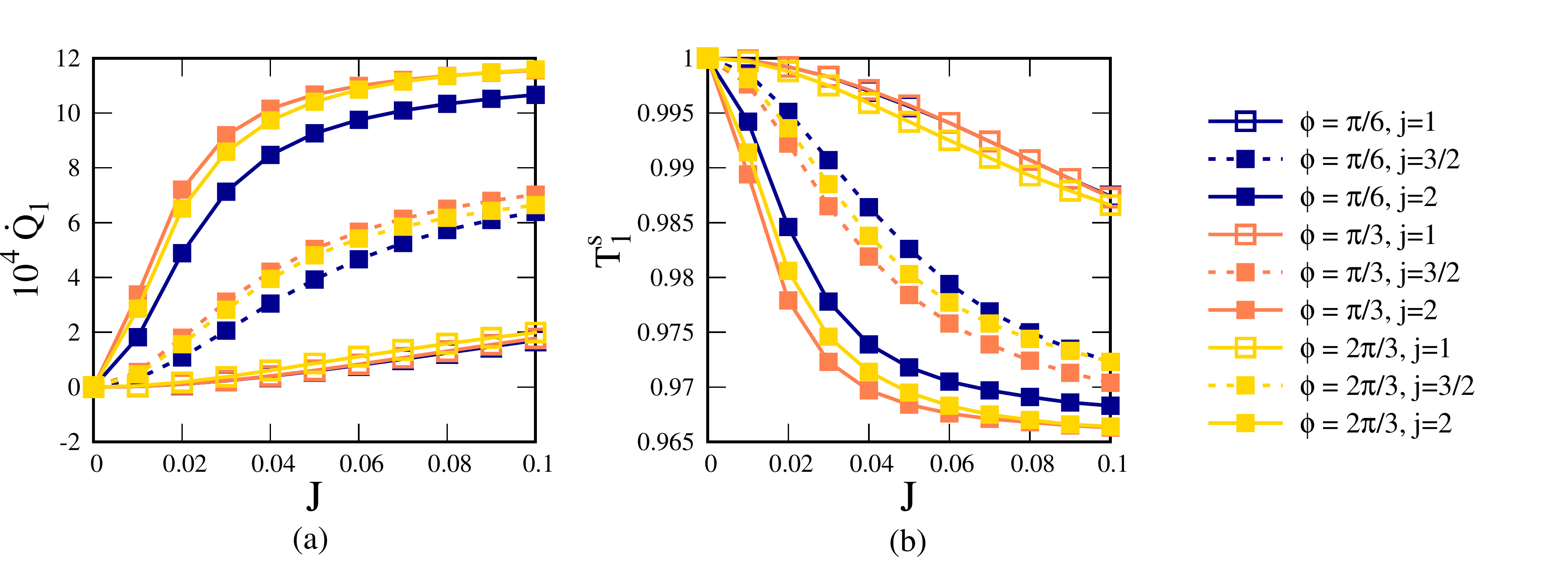}
    \caption{(Color online) \textbf{Trends of (a) heat current  and (b) local temperature in the steady state  (vertical axis) against the interaction strength, $J$ for a two-spin BB Hamiltonian (horizontal axis)}.
    Hollow squares with solid line, solid squares with dashed line and solid squares with solid line respectively represent \(j=1\), \(3/2\) and \(2\) respectively while blue, orange and yellow respectively denotes the phases, \(\phi =\pi/6\), \(\pi/3\) and \(2 \pi/3\).
    All other specifications are same as Fig. \ref{fig:xyz_two_spin_local}. All the axes are dimensionless.
    }
    \label{fig:bbh_two_spin_local}
\end{figure*}

\begin{figure}
    \centering
    \includegraphics[width=\linewidth]{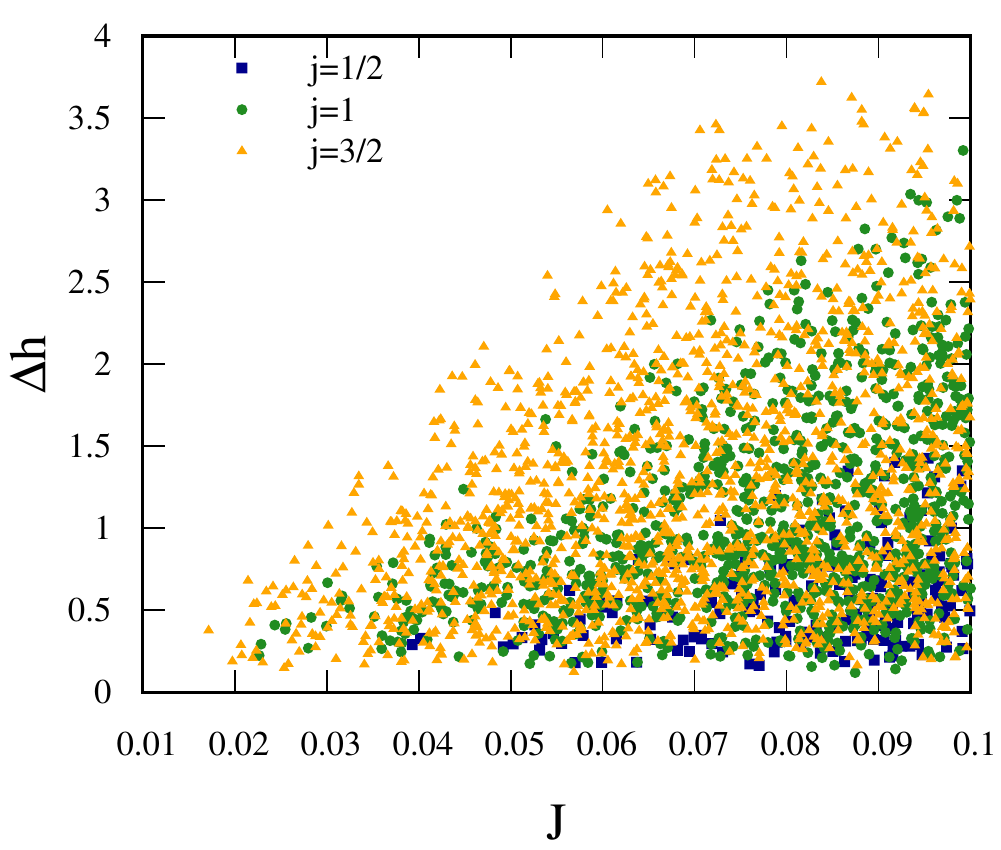}
    \caption{(Color online.) \textbf{Scattered plot of the steady-state temperature for two-spin refrigerators only when cooling  occurs in the $(J$, $\Delta h = h_2 - h_1)$-plane}. Squares, circles and triangles represent respectively $j=1/2$, $1$ and $3/2$. 
    Other specifications are same as Fig. \ref{fig:xyz_two_spin_local}. Among $7 \times 10^3$ choices of parameters of the refrigerator based on the $XX$ model, $2.7\%$, $14.3\%$ and $23.82\%$ of situations are found to exhibit cooling in the steady state with $j=1/2$, $j=1$ and $j=3/2$ respectively.  Both the axes are dimensionless.}
    \label{fig:scatter}
\end{figure}

\section{Two-spin quantum refrigerators}
\label{sec:twospinR}

We now discuss the performance of quantum refrigerators built with two spins, where one of the spins is cooled and the other spin works as the refrigerator. Unless otherwise mentioned, in the rest of the paper, we always choose the first spin, i.e.,  $r=1$ to be the target spin for cooling.

\subsection{System of two identical spins}
\label{subsec:like_spins}

Let us consider two identical interacting spin-$j$ particles constituting the system, and increase the value of $j$ simultaneously for both the spins to study how the refrigeration of one of the spins depends on $j$. Unless otherwise mentioned, in all our analysis, we use the trace distance to define the DLT. 
For computing heat current, local temperature and entropy, we solve the local as well as global QME using the Runge-Kutta fourth order technique, and determine the reduced state of the spin-$j$ particle in the steady state that is used to compute the relevant quantities.

\subsection*{Tuning refrigerator with system parameters} 

\textbf{\(XYZ\) model as refrigerator. } We first  consider the $XYZ$-type interaction between the spins, and solve the time-dependent state of the system using the local QME to compute $\dot{Q}_1$, $T_1^{s}$, and $S_N^s$ corresponding to the first spin. Fig.~\ref{fig:xyz_two_spin_local} depicts the variations of these quantities as a function of $j$, clearly indicating significant advantage in cooling of the first spin when the dimension of the Hilbert spaces corresponding to the spins increases. For example, with \(J=0.09\), in case of spin-$1/2$, the decrease in temperature at the steady state from the initial state is \(\approx 0.59\%\) while it is \(3.1\%\) for the refrigerator with spin-$4$ systems.
Also, our results indicate that a higher value of $J$ favours the cooling of the first spin, compared to a lower value. In our analysis, we have kept the value of $J$ to be in such a range that the local QME can be applied. However, the improvement in cooling for higher values of $J$ indicates the need for an investigation with the global QME, which we shall discuss in the subsequent subsections.

Note that the results presented in Fig.~\ref{fig:xyz_two_spin_local} is for the  case of $\gamma=0$ and $\Delta=0$, representing the \(XX\) Hamiltonian. Our data suggests that even in the presence of the $xy$- and the $z$-anisotropy in the interaction, the dimensional advantage in cooling persists although  the variation of the relevant quantities is almost negligible with non-zero values of the anisotropies for a fixed value of $j$, especially with low $j$ values.  For  high $j$, the slight change in the local temperature happens with the introduction of \(\gamma\) and \(\Delta\) and the results suggest that the performance of the refrigerator based on the $XXZ$ model is the  best among the class of $XYZ$ models. Therefore, the behaviors of the heat current, entropy and local temperature depicted in Fig~\ref{fig:xyz_two_spin_local} faithfully capture all the relevant information regarding the effect of increasing spin-dimension and the spin-spin interaction strength on performance of the refrigerator. Note that the positive or negative  coupling strength, \(J\), leads to the same local temperature in the steady state.

Let us check whether  the local master equation in this situation is thermodynamically valid (see Sec.~\ref{subsec:local_heat_current}). For cooling qudit $1$ in a two-qudit system, it extracts heat from the environment, and one would expect~\cite{dechiara2018}
\begin{eqnarray}
\dot{Q}_1>0,  \dot{Q}_2<0, \,  \mbox{and} \, \dot{W}>0,
\label{eq:thermodynamic_validity}
\end{eqnarray}
which are consistent and valid in thermodynamical processes, where $\dot{W}=\mbox{Tr}[\{\sum_r\mathcal{L}_r(\rho_s)\}H_{int}]$. In Fig.~\ref{fig:xyz_two_spin_validity}, we plot $\dot{Q}_r$, $\dot{W}$,  $\dot{\Sigma}$ (see Sec.~\ref{subsec:local_heat_current} for definition), and $T^0_1-T^s_1$ with varying initial magnetic field of the first system in the case of local master equation for systems of two spin-$1/2$, and two spin-$1$ particles,  exhibiting agreement with Eq.~(\ref{eq:thermodynamic_validity}). This is true for all situations presented in this paper where the local master equation is applied, implying thermodynamic consistency. Note, further, that in all calculations, the indication of cooling of qudit $1$ by the distance-based definition of temperature (i.e., $T_1^0-T_1^s$) is used to check the validity of Eq.~(\ref{eq:thermodynamic_validity}). This puts the notion of refrigeration by defining the distance-based temperature on stronger footing.

\textbf{Refrigerator with bilinear-biquadratic interactions.} Using the local approach, we also investigate the performance of the two-spin refrigerator when the spin-spin interactions are governed by the BB Hamiltonian (see Fig.~\ref{fig:bbh_two_spin_local}), and have found the results to be qualitatively similar to that reported in Fig.~\ref{fig:xyz_two_spin_local}. The dimensional advantage of cooling is  present irrespective of the phase of the system from which the spin-spin interaction parameter is chosen. Specifically,  the parameters chosen for demonstration reveals that the minimum temperature is obtained when  the corresponding system at equilibrium belongs to the critical phase. 

\noindent\emph{Note.} A comment on the choice of the system parameters for the demonstration of refrigeration is in order here. Although numerous points in the parameters space of the system parameters exists where a local steady-state cooling for the first spin is observed, the total volume of the parameter space that represents such refrigerators is small compared to the entire parameter space although it increases with the increase of the spin-dimension. In Fig.~\ref{fig:scatter}, we depict, for $j=\frac{1}{2},1,\frac{3}{2}$, the points in the parameter space of $J$ and $\Delta h=h_2-h_1$ for which a steady-state cooling of at least $T_1^0-T_1^s=10^{-3}$ is obtained. The fraction of points representing a refrigerator increases with an increase in $j$ $2.7\%$, $14.3\%$ and $23.82\%$ for \(j = 1/2, \, 1, \,\text{and}\, 3/2\) respectively, demonstrating again a dimensional advantage in the accessibility of parameter space in building a quantum refrigerator. Note also the higher clustering of the accessible points in parameter space towards a high value of $J$, indicating the need for performing a global QME-based analysis of the system. Interestingly, however, we notice that there exists  a forbidden regime in the \((J, \Delta h)\)-plane where the cooling with  $XX$ interactions does not occur and it decreases with dimensions.

\begin{figure}
    \centering
    \includegraphics[width=\linewidth]{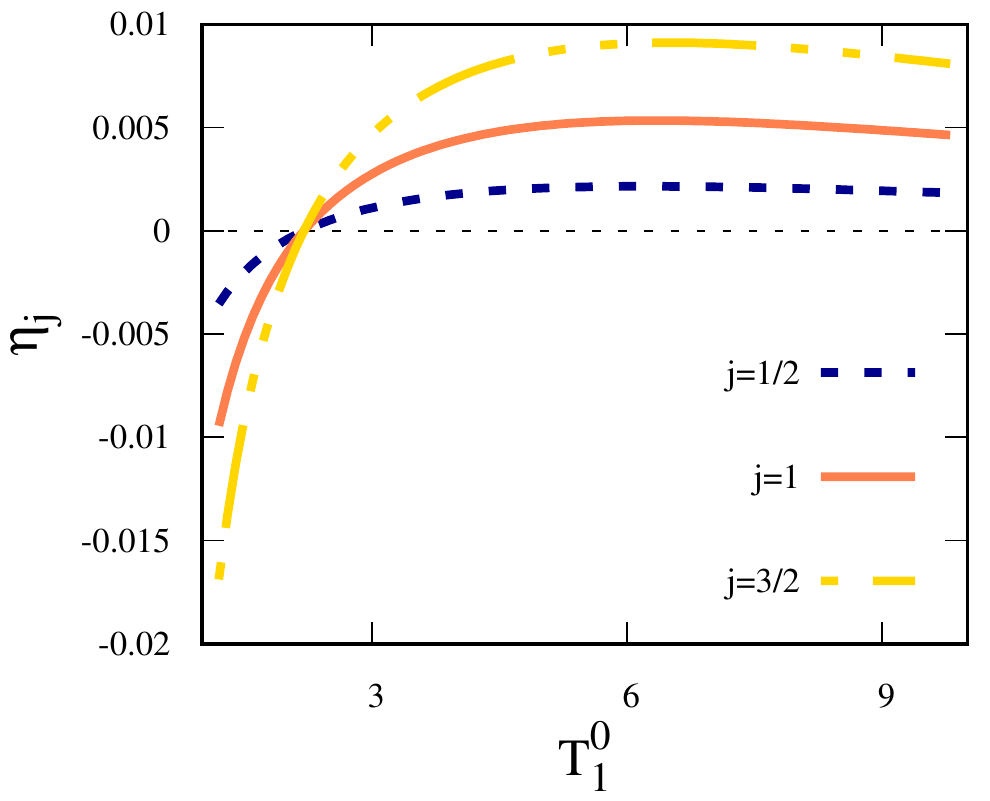}
    \caption{(Color online) \textbf{Dependence of steady state cooling factor, \(\eta_j\) (vertical axis) vs. initial temperature (horizontal axis) for two-spin refrigerator}. Dashed, solid  and dotted lines represent spin quantum number, $j = 1/2$ $1$ and $3/2$ respectively. Here $J = 0.05$ and $T_2^0-T_1^0 =0.4$. Other specifications are same as in Fig. \ref{fig:xyz_two_spin_local}. Both the axes are dimensionless. }
    \label{fig:temperature_tuning}
\end{figure}

\subsection*{Tuning refrigerator with bath temperature} 

Along with system parameters, it is also important to investigate how the performance of the refrigerator can be controlled when one has access to the tunable parameters of the thermal baths, such as the bath temperatures $T_r^0$. Towards this aim, we define a steady-state cooling factor relative to the initial temperature of the cold spin-$j$ particle in the system, as 
\begin{eqnarray}
\eta_j=\frac{T_1^0-T_1^s}{T_1^0}. 
\end{eqnarray} 
In Fig.~\ref{fig:temperature_tuning}, we plot the variation of $\eta_j$ as function of $T_1^0$, which exhibits a critical point $T_1^c$ on the $T_1^0$-axis corresponding to a zero-crossing of $\eta_j$. For $T_1^0<T_1^c$, a steady-state heating of the first spin takes place represented by a negative value of $\eta_j$, while when $T_1^0>T_1^c$, a positive value of $\eta_j$ is obtained due to the occurrence of a steady-state cooling. Note that for the reported data, the critical point $T_1^c$ corresponds to the situation described in Proposition I, such that 
\begin{eqnarray}
\frac{h_1}{T_1^c}=\frac{h_2}{T_2^0},
\end{eqnarray}
ensuring that no evolution of the system takes place. Note also that our numerical analysis clearly suggests that 
\begin{eqnarray}
\eta_{j=\frac{1}{2}} < \eta_{j=1} < \eta_{j=\frac{3}{2}},
\end{eqnarray}
thereby exhibiting the importance of higher dimensional subsystems in enhancing the performance of the designed refrigerator.

\begin{figure}
    \centering
    \includegraphics[width=\linewidth]{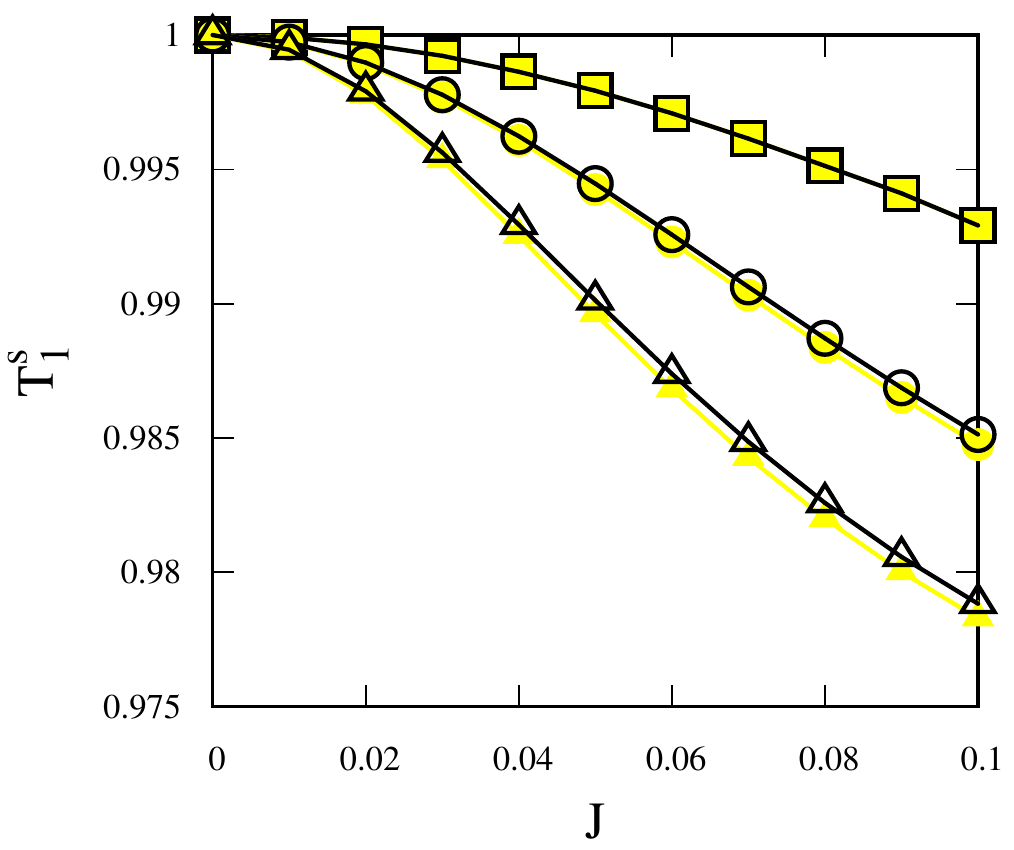}
    \caption{(Color online) Behavior of steady-state local temperature of the target spin obtained via trace  and relative entropy distance  (ordinate) with respect to interaction strength, $J$ (abscissa) for a two-spin system. Solid and hollow symbols represent trace  and  relative entropy distance. Squares, circles and triangles are respectively for refrigerators with two identical spins having \(j=1/2,\, 1, \, 3/2\). Other specifications are same as Fig. \ref{fig:xyz_two_spin_local}. All the axes are dimensionless. }
    \label{fig:different_distance_measure}
\end{figure}

\begin{figure}
    \centering
    \includegraphics[width=\linewidth]{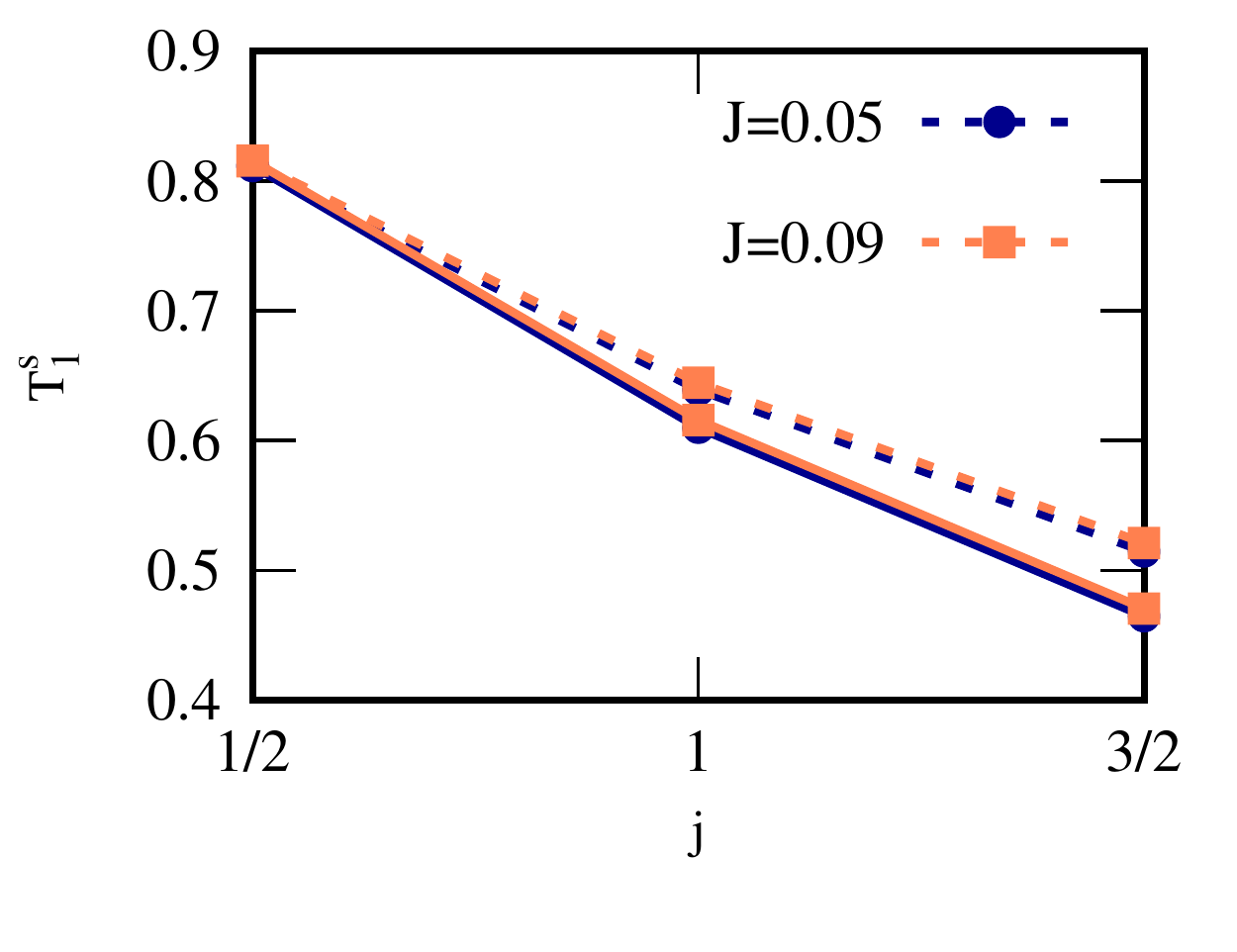}
    \caption{(Color online) \textbf{Steady-state local temperature  (ordinate) with the variation of spin quantum number $j$ (abscissa).} The dashed  line is obtained by solving global QME for the $XXZ$ model for a two-spin refrigerator with spin-$j$ systems and the solid line is for the refrigerator consisting of a spin-$1/2$ and spin-$j$ particles. 
    Here $J=0.05$ (circles) and $J=0.09$ (squares) with $J\Delta=-1.0$ while the strength of the magnetic fields  and the spin-bath interactions are chosen as $h_1=1.1$, $h_2=1.3$, $\alpha_1,\alpha_2=10^{-3}$ and $\omega_c=10^3$ respectively,  and the initial temperatures of each spin are same as in Fig. \ref{fig:xyz_two_spin_local}. See text for the improvement obtained with global QME over the local ones.  Both the axes are dimensionless. 
    }
    \label{fig:global_two_spins}
\end{figure}

\begin{figure*}
    \centering
    \includegraphics[width=\textwidth]{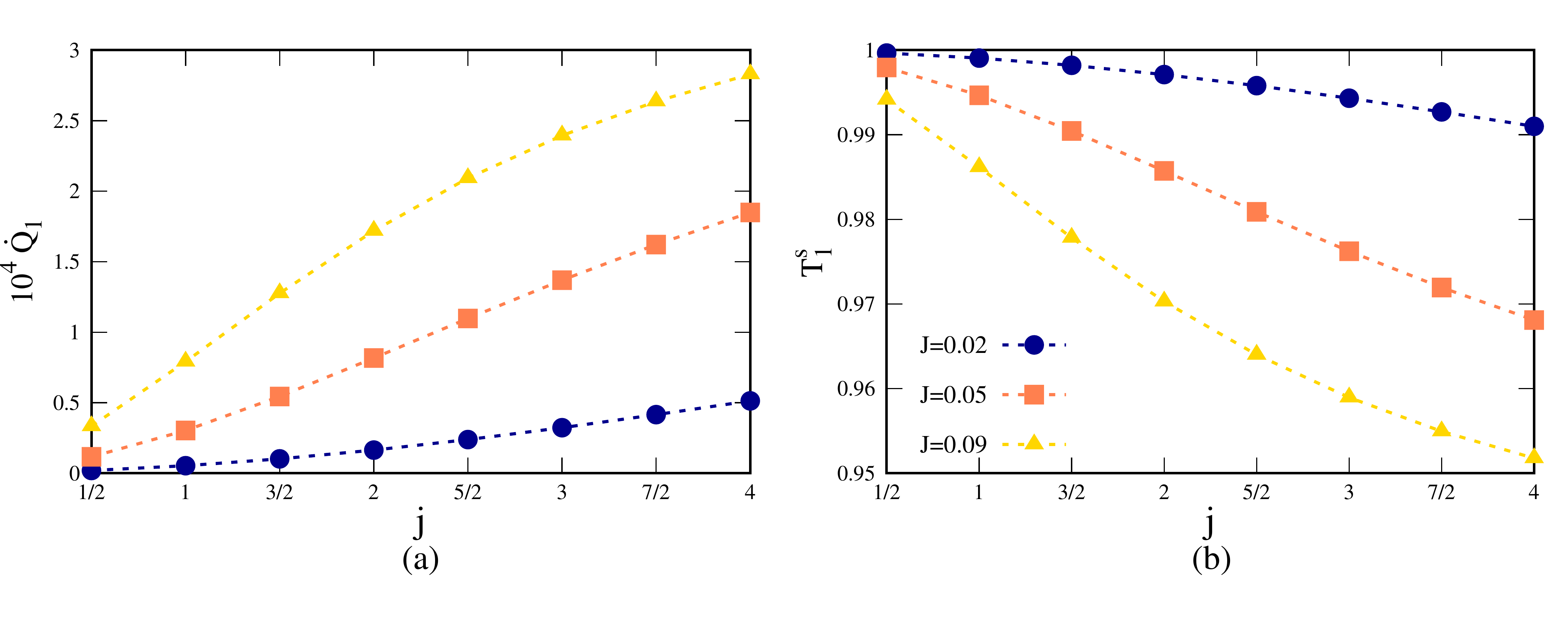}
    \caption{(Color online) (a) \(\dot{Q_1}\) and (b) \(T_1^s\)  (ordinate) against the dimension, \(j\) (abscissa). Here the refrigerator is built with a spin-$1/2$ and a spin-$j$ particles governed by the Hamiltonian \(H_{xx}\) in Eq. (\ref{spin1/2_spinj_hamiltonian}).  All other specifications are same as in Fig. \ref{fig:xyz_two_spin_local}.  All the axes are dimensionless.}
    \label{fig:twospinnonidentical}
\end{figure*}

\textbf{Local temperature with different distance measures.} At this point, it is natural to ask whether the reported results remain invariant under a change in the choice of the distance measure used to quantify the DLT. We answer this question affirmatively. Fig.~\ref{fig:different_distance_measure} depicts a comparison between the DLT values obtained by using the trace distance and the relative entropy distance, defined as
 \(   S(\sigma_{1} || \sigma_{2}) = \text{Tr}\left[\sigma_{1} \log_2 \sigma_{1} -\sigma_{1} \log_2 \sigma_{2}\right]\),
for two density matrices $\sigma_1$ and $\sigma_2$. 
While the two measures provide identical results for qubit systems, the values of the DLTs differ by $\sim10^{-3}$ with increasing $j$. Nonetheless, the qualitative behavior remains similar in all these situations. It is also noteworthy that the difference is very small for low values of the spin-spin interaction strength, and increases very slowly with an increase in $J$. We also check the performance of the DLT using Uhlmann fidelity~\cite{uhlmann1976} as the distance measure, which coincides with the DLT using relative entropy distance.

\subsection*{Refrigeration using the global QME}

A question that naturally arises is to whether the results corresponding to a quantum refrigerator obtained using the local QME remains the same even in situations where a global QME is appropriate to describe the dynamics of the system. To answer this question, we find that in the case of a two-spin models described by the Hamiltonian $H_{xyz}$, cooling of the first spin takes place only with a non-zero value of $\Delta J$ (see Fig.~\ref{fig:global_two_spins} for a typical cooling phenomena for the first spin). This is in stark contrast with the situations discussed so far involving the local QME, where the $zz$-interaction term in $H_{xyz}$ does not have any significant effect on cooling (cf.~\cite{konar2021}). However, even in the case of the global QME, the features like significant dimensional advantage remains unaltered, and the amount of refrigeration of the first spin is much higher in comparison to the case of the local QME. 
For example, for spin-1/2 systems, the percentage of cooling in the first spin is approximately \(18.8\%\) with global QME   while for spin-$3/2$ quantum refrigerator, it is  \(53.5\%\) for the $XXZ$ model refrigerator with \(J=0.05\) and \(J \Delta  =-1\).

\subsection{System of two different spins}
\label{subsec:mixed_spin}

 \begin{figure*}
    \centering
    \includegraphics[width=\textwidth]{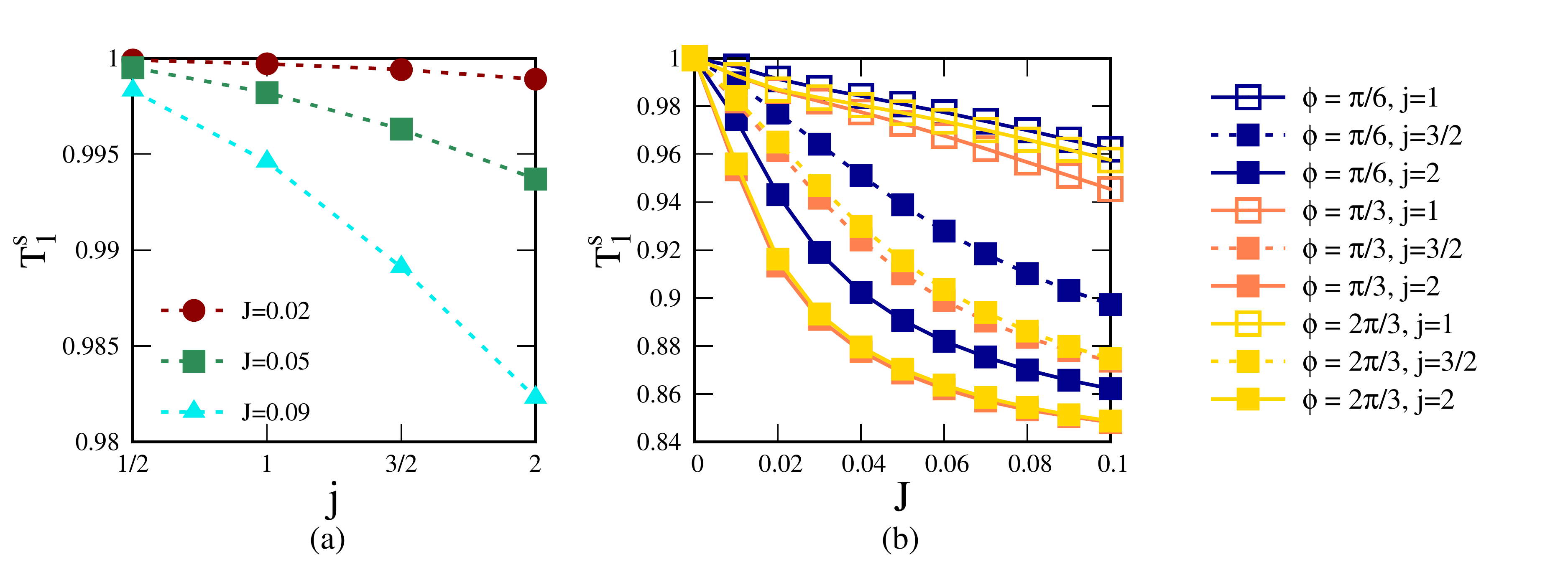}
   \caption{(Color online) (a) Local temperature (vertical axis) as a function of the dimension of  spins $j$ (horizontal axis) with $N = 3$ (three-spin refrigerator) for Heisenberg $XX$ Hamiltonian. Here we choose $h_1=1.5$, $h_2=2.5$, $h_3=3.5$ and $\Gamma=0.05$,  and the initial temperature of each spin is $T_1(0)=1$, $T_2(0)=1.1$ and $T_3(0)=1.5$.
   (b) \(T_1^s\) (ordinate) against interaction strength \(J\) (abscissa) with \(N=3\) for the refrigerator based on the BB Hamiltonian for different \(\phi\) values. Initial values of the magnetic fields, temperature and bath-system interaction strengths are same as in the $XX$ model. 
   All the axes are dimensionless.}
    \label{fig:xx_bbh}
\end{figure*}

Let us design a refrigerator with two spins  having different values of $j$, and focus specifically on the situation where $j=\frac{1}{2}$ for the first spin $(r=1)$, while for the second spin $(r=2)$, $j$ can take an arbitrary value.  While it is known that a qubit can be cooled in a qubit-qutrit system with specific interaction between them~\cite{linden2010}, it is not yet clear whether increasing the Hilbert space dimension of the second party in a $2\times (2j+1)$ system provides any advantage to the refrigeration of the qubit system. To address this question, we consider the Hamiltonian modeling the interaction between the spin-$1/2$ and spin-$j$ particle to be
\begin{equation}
    H_{xx}=J[\tilde{S}_1^xS_2^x+\tilde{S}_1^yS_2^y],
    \label{spin1/2_spinj_hamiltonian}
\end{equation}
where $\tilde{S}$ ($S$) represents the spin operator corresponding to the spin-$\frac{1}{2}$ (spin-$j$) subsystem, and $J$ is the strength of the spin-spin interaction.  In Figs. \ref{fig:twospinnonidentical}(a) and (b), we respectively observe the patterns of $\dot{Q}_1$ and $T_1^{s}$ of the spin-$\frac{1}{2}$ particle  by varying $j$ for the second spin. With an increasing $j$ for the second spin, $\dot{Q}_1$ ($T_1^{s}$) starts from a low positive value and then increases (starts from a value $\approx T_1^0$ and then decreases), exhibiting again the dimensional advantage in cooling the first spin. Surprisingly, we observe that in this non-identical scenario, the minimum temperature   corresponding  to \(j=4\) of the second spin (i.e., the decrease of temperature, \( \approx 4.82\%\)) is much lower (higher) than that obtained for the scenario with identical spins (the decrease from the initial temperature, \(\approx 3.1\%\) (comparing Figs. \ref{fig:xyz_two_spin_local} and \ref{fig:twospinnonidentical}).  
We also perform the same analysis using the global QME to find a more pronounced dimensional advantage. Specifically, with the $XXZ$ model (\(J=0.05, J \Delta = -1 \)), we find that a $18.9\%$ cooling of the first spin in the case of $j=\frac{1}{2}$ for the second spin occurs while  it becomes  $48.53\%$, when the spin quantum number of the second spin is increased to $j=3/2$.

\section{Refrigeration in three-spin systems}
\label{sec:threespinR}

Let us now move to a set up of refrigerator consisting of three identical spin-$j$ particles, each of which is connected to a local thermal bath as shown in Fig. \ref{fig:schematic}.
Starting with the product state of a local Hamiltonian, \(H_{loc}\), the system evolves according to Hamiltonian, \(H_{xyz}\) or \(H_{B}\) at \(t>0\). In case of the $XYZ$ refrigerator, we consider an isotropic case ($\gamma=0,\Delta=0$) for demonstration of the performance of the refrigerator in a local QME approach. Fig.~\ref{fig:xx_bbh} depicts the variations of $\dot{Q}_1$, $T_1^{s}$, and $S_N^s$ as functions of $j$ for different values of $J$, clearly demonstrating a dimensional advantage. Note that although the qualitative results on refrigeration of the first spin using the three-spin system remains similar to its two-spin variant (see Sec.~\ref{sec:twospinR}), quantitatively the two-spin refrigerator performs better than the three-spin one, which can be seen by comparing Fig.~\ref{fig:xx_bbh} with Fig.~\ref{fig:xyz_two_spin_local}.

In case of refrigerator with three-spins governed by the BB Hamiltonian, $T_1^{s}$ and $S_N^s$  again exhibit increasing  cooling of the first spin with increasing $J$ as well as with the increase of spin quantum number, \(j\).  As observed for  the two-spin refrigerator, the phase dependence also remains unaltered. However, the heat current,  $\dot{Q}_1$ exhibits a non-monotonic variation with $J$ for $j=2$  when $\phi = \pi/3$ and $2\pi/3$, and becomes negative for moderate and high values of $J$.  We point out here that we have defined the local heat current following the global approach (see Sec.~\ref{subsec:osd}) which may lead to such anomalous behavior in the heat current (cf. \cite{Wichterich2007, barra2015, strasberg2017,De_Chiara2018,ghoshal2021,konar2021}).

\section{Conclusion}
\label{sec:conclu}

Summarizing, we have designed a quantum refrigerator built of a few  spins whose individual Hilbert space dimensions can go beyond the qubits, or qutrits. The spins are considered to be interacting among each other via the $XYZ$  and  the bilinear biquadratic interactions, while  each of the spins are locally interacting with the bosonic baths. So far, such machines have  typically been built with spin-$1/2$ or spin-$1$ systems, and the quantifiers of the performances of these machines, such as definitions of local temperature for the constituent subsystems,  are designed accordingly. To deal with the higher dimensional systems, in this paper, we propose a new definition of local  temperature based on the minimum distance between the dynamical state of a spin-$j$ particle in the steady state, and a canonical thermal state of the same particle, which proves to be a faithful quantifier for the performance of the designed refrigerator. The definition is proved to be consistent with the existing definitions for qubit systems, and the  behavior of the distance-based local temperature is found to be in agreement with the local heat current and the entropy of the subsystems. Furthermore, our results can also shown to be consistent with the entropy product rate, flows of heat current between system and environment and  the work rate which are required to have a valid thermodynamic process. 
We observed that our setup leads to a cooling of one of the spins in the system, which enhances with the increase of the spin quantum number of the spins, and thereby with the increase of the Hilbert space dimension.  Hence it  establishes the dimensional advantage in the refrigerators, in the same vein as, for example, in~\cite{silva2016} (cf.~\cite{clivaz2019} for a different setup where spin-$1/2$ systems are found to be most advantageous).  On our way to  verify these results by using both local and global quantum master equations, we have also analytically derived the form of the Lindblad operators corresponding to the individual spins while constructing the dissipator for the local quantum master equation.

Miniaturisation of devices are necessary to fulfil  the current way of living. Although most of these devices work according to the laws of classical physics, they have now started knocking at the door of the quantum world due to immense advancement in the design and control of machines at the microscopic scale. In recent years,  it has been established that appliances  based on quantum mechanics can remarkably enhance the efficiencies compared to that obtained from  the existing ones, thereby revolutionizing the world of technologies. In this respect, our work explores and manifests building small quantum refrigerators using quantum spin systems with large spins. The scope for future exploration from our work is immense. For instance, note that starting from a microscopic quantum thermal machine, there exists two routes to macroscopicity -- (a) by increasing dimension of individual subsystems of a composite quantum system while keeping the number of subsystems small, and (b) by having a large number of small subsystems~\cite{mohammady2018, arisoy2021}. Our results explores the former, while the latter also gained some interests in the recent past~\cite{george2011,deniz2017,lekscha2018, kloc2019,hong2020,revathy2020}. It will be interesting to find out the hierarchies, if any,  among the macroscopic devices obtained following these two different routes.

\acknowledgements

TKK, SG and ASD acknowledge the support from the Interdisciplinary Cyber Physical Systems (ICPS) program of the Department of Science and Technology (DST), India, Grant No.: DST/ICPS/QuST/Theme- 1/2019/23. AKP acknowledges the Seed Grant from IIT Palakkad. We  acknowledge the use of \href{https://github.com/titaschanda/QIClib}{QIClib} -- a modern C++ library for general purpose quantum information processing and quantum computing (\url{https://titaschanda.github.io/QIClib}), and the cluster computing facility at the Harish-Chandra Research Institute. This research was supported in part by the INFOSYS scholarship for senior students.

\appendix
\section{Proposition I for two-qubit systems}
\label{prop_xx}

We now elaborate the proof of Proposition 1. We explicitly show that $[H_{sys},\rho_0]=0$ when the condition $\frac{h_1}{T_1^0}=\frac{h_2}{T_2^0}$ is satisfied for a two-qubit system described by the XX Hamiltonian. This calculation, however, can be generalized for arbitrary number of qudits as long as the system Hamiltonian is of XX type.

Let us label the two qubits as $1$ and $2$. The initial states of the qubits $1$ and $2$ are given by  $\rho_1=\tau\dyad{0}{0}+(1-\tau)\dyad{1}{1}$ and $\rho_2=\tau'\dyad{0}{0}+(1-\tau')\dyad{1}{1}$ respectively, where $\tau=e^{-h_1/T_1}/Z_1$, $\tau'=e^{-h_2/T_2}/Z_2$ and $Z_i=e^{-h_i/T_i}+e^{h_i/T_i}$. The initial state of the total system is diagonal in the computational basis, given by 
\begin{eqnarray}
\rho_0&=&\rho_1\otimes\rho_2\nonumber\\
&=&\tau\tau'\dyad{00}{00}+\tau(1-\tau')\dyad{01}{01}+\tau'(1-\tau)\dyad{10}{10}\nonumber\\
&&+(1-\tau)(1-\tau')\dyad{11}{11}.
\end{eqnarray}
The total Hamiltonian of the system is given by the form $H_{sys}=H_{loc}+H_{int}$. Note that the local part of the Hamiltonian is also diagonal in the computational basis, and, therefore, commutes with the initial state. On the other hand,  $H_{int}$ consists of off-diagonal elements, and can be written in computational basis as
\begin{equation}
    H_{int}=J(\dyad{01}{10}+\dyad{10}{01}). 
\end{equation}
Now one can evaluate 
\begin{widetext}
\begin{eqnarray}
[H_{sys},\rho_0]&=&[H_{int},\rho_0]\nonumber\\&=&[J(\dyad{01}{10}+\dyad{10}{01}),\tau(1-\tau')\dyad{01}{01}+\tau'(1-\tau)\dyad{10}{10}]\nonumber\\&=&-J\tau(1-\tau')\dyad{01}{10}+J\tau'(1-\tau)\dyad{01}{10}+J\tau(1-\tau')\dyad{10}{01}-J\tau'(1-\tau)\dyad{10}{01}\nonumber\\&=&J(\tau-\tau')[\dyad{10}{01}-\dyad{01}{10}]. 
\end{eqnarray}
\end{widetext}
Evidently, $\tau=\tau'$ implies  $[H_{sys},\rho_0]=0$. Explicitly writing $\tau$ and $\tau^\prime$ in terms of $h_i$ and $T_i^0$, one obtains $\frac{h_1}{T_1}=\frac{h_2}{T_2}$.

\bibliographystyle{apsrev4-1}
\bibliography{ref}

\end{document}